\documentclass{resonance}

\usepackage{times}
\usepackage{graphicx}
\usepackage{fancybox}
\usepackage{amsmath}
\usepackage{amssymb}
\usepackage{float}

     {\begin{Sbox}\begin{minipage}}%
     {\end{minipage}\end{Sbox}\fbox{\TheSbox}}

\begin{document}

\title{Compact objects and Black Holes: Nobel Prize for Physics 2020} 

\author{J. S. Bagla}

\maketitle

\authorIntro{\includegraphics[width=2cm]{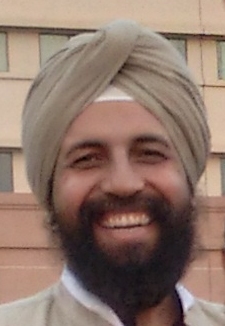}\\
Jasjeet works at IISER Mohali.  He is interested in diverse problems
in physics.  His research is in cosmology and galaxy formation.}

\begin{abstract}
  The Nobel Prize in Physics 2020 has been divided, one half awarded to
  Roger Penrose {\sl for the discovery that black hole formation is a
    robust prediction of the general theory of relativity.} and the
  other half jointly to Reinhard Genzel and Andrea Ghez {\sl for the
    discovery of a supermassive compact object at the centre of our
    galaxy.}  Here we describe their work and put it in historical
  context and discuss specific advances that have been rewarded.
\end{abstract}

\monthyear{month year}
\artNature{GENERAL ARTICLE}

\section{Dark Stars}

The Nobel prize this year has been awarded for research that
demonstrated that black holes can form in realistic astrophysical
situations, and observational evidence for a supermassive compact
object at the centre of our Galaxy.
Black holes are exotic objects.
These have been in popular imagination and these have also been used
in movies.

Black holes are thought of as objects from where nothing can escape.
The first speculation about such objects, then named as {\sl dark
  stars} was the result of combining an increasing confidence in Newtonian
gravity and the speed of light as determined by Ole
R\"omer\mfnote{R\"omer used the eclipses of Jupiter's satellites to
  measure the speed of light.  He found that the estimated time
  between successive eclipses is shorter when the Earth is moving
  towards Jupiter and longer when the Earth is moving away from
  Jupiter.  He correctly interpreted this as a result of the finite
  speed of light.}.
The speed of light is many orders of magnitude larger than the speeds
one encounters in day to day life.
In 18th century, John Michel and Pierre-Simon de Laplace speculated
about existence of stars so massive that light emitted cannot escape
to infinity. 
A simple calculation shows that the mass and the radius of such stars
satisfies the relation based on the standard formula for escape velocity: 
\begin{equation}
  R \leq \frac{2 G M}{c^2}
\end{equation}
Here, $G$ is the universal constant of gravitation, $M$ is the mass of
the star, $c$ is the speed of light and $R$ is the radius\mfnote{For
  masses that are familiar to us, the radius is very small.  The
  radius corresponding to the Sun is just under $3$~km, whereas the
  radius of the Sun is close to $7\times 10^5$~km.}.
It is to be noted though that in this scenario, any observer at a
finite distance from the star will still see the light from the star.
Michel even speculated that {\sl the only way to discover the presence
  of such a dark star may be to study motion of luminous objects in
  its vicinity. }
Laplace conjectured that a significant fraction of objects in the
Universe may be dark stars.

\section{General Relativity and Black Holes}

Albert Einstein introduced the special theory of relativity in 1905
and this elevated the speed of light from a very high speed to the
maximum speed possible in nature.
General theory of relativity, introduced by Einstein in 1915,
connected the curvature of space-time with the notion of gravity while
retaining aspects of the special theory.

Karl Schwarzschild\mfnote{Schwarzschild presented this solution in
  1916. Unfortunately he died  shortly after this.} presented the
first exact solution for the equations of the general theory of
relativity. 
He presented the solution for gravitational field of a point mass.
Two aspects of this solution are noteworthy: the gravitational field
is singular with a singularity at the position of the point mass, and,
nothing can escape from a sphere of radius given by Eqn.(1).
This sphere acts as a one way membrane and is referred to as the
horizon\mfnote{The understanding that this is a one way membrane came
  much later with contributions from many, including Roger Penrose.
  At the time, it was believed that there is a singularity at the
  surface of this sphere but it was eventually understood that this
  problem is due to the choice of coordinates in the Schwarzschild
  metric.}.   
The radius of this sphere is referred to as the Schwarzschild radius.  
This was the first relativistic expression of what we now call a black
hole. 

It was found that at large distances from the black hole, the
gravitational field and orbits deduced from Schwarzschild metric are
well approximated by Newtonian gravity with small corrections.
These corrections lead to precession of bound orbits: precession of
Mercury's orbit was one of the first verifications of the general
theory of relativity. 
Close to the black hole the differences between the two are very
significant.
There are no stable bound orbits possible with an approach radius
smaller than three times the Schwarzschild radius.
Thus any massive particle approaching the black hole closer than this
distance is expected to fall into the black hole.
Photons can orbit around such a black hole with an orbital radius
equal to $1.5$ times the Schwarzschild radius.
Photons coming in from larger radii and approaching closer than this
distance fall into the black hole.
At large distances, photons get deflected from straight line by a
small amount.
This was observationally verified for the first time by Dyson,
Eddington and Davidson (1920) during the total solar eclipse on May
29, 1919.
This observational verification was critical in making general theory
of relativity as the accepted theory of gravitation and international
fame for Einstein. 

Schwarzschild also presented a solution for space-time due to a star
and he showed that this is not singular and there is no horizon in
such a case.
He assumed the star to have a finite radius and uniform density. 
In light of this, the point mass solution remained a curiosity for
some time.

Chandrasekhar limit for white dwarf stars\mfnote{S. Chandrasekhar
  derived the limit by combining special relativity with quantum
  statistics.  He showed that a consequence of this combination is
  that as a more massive white dwarf star has electrons that are
  relativistic and the effective equation of state leads to an
  instability if a white dwarf star made predominantly of Helium has a
mass more than $1.4$~M$_\odot$, where M$_\odot = 2 \times 10^{30}$~kg
is the mass of the Sun.} raised the question of what
happens if such a star goes beyond the mass limit. 

In a star like the Sun, gravitational pull is finely balanced by a
combination of gas pressure and radiation pressure.
Thus one requires a time dependent study in order to address the
question raised by Chandrasekhar's computation of the mass limit for
white dwarf stars.

\section{Collapse and Singularities}

Bishveshwar Datt from Presidency College, Kolkata solved for time
dependent evolution of spherically symmetric density distributions.
His interest was in cosmological expansion of inhomogeneous regions.
He published his solutions for the cosmological scenario and an
expanding universe in 1938.
He passed away in the same year during a surgery.
His paper from 1938 was republished in 1997 as a golden oldie in the
journal {\sl General Relativity and Gravitation}. 
The same solution was discovered independently by Oppenheimer and
Snyder in 1939 and applied to a collapsing star.
They showed that in absence of pressure the star collapses and
continues to collapse.
However, the collapse slows down from the perspective of a distant
observer as the radius of the star approaches the Schwarzschild
radius.  

A very important link in the development of ideas about time dependent
space-times was provided by Professor Amal Kumar Raychaudhuri.
Working in Ashutosh College in Kolkata, he developed a general
equation for describing local evolution of such space-times without
imposing any restrictions of symmetry or constraining the type of
matter that drives the evolution of space-time.
The equation describes how matter moves in such a space time and how
the space time evolves with it. 
The equation, named the Raychaudhuri equation in his honour describes
space-time in terms of an overall expansion or contraction, rotation,
and shear\cite{2008Res13319K}.
However, the assumption of zero pressure raised doubts about the
relevance of the solution for real stars.

We have discussed above that the Schwarzschild solution has a
singularity at the centre and it has a horizon enclosing the
singularity.
The development of this understanding came through the work of Roger
Penrose. 
If we consider the view of a distant observer, gravitational
time dilation implies that it takes an infinite time for any object to
fall into the horizon.
Penrose introduced a set of coordinates that demonstrate that in the
frame of a particle falling into the black hole, fall towards the
singularity is inevitable once it crosses the horizon and it happens
in finite time. 
More interestingly, if this particle is emitting light then all light
falls into the black hole once the particle crosses the horizon.
Penrose attributed the idea to Eddington and Finkelstein and these
coordinates are known as Eddington-Finkelstein coordinates.
Thus Penrose was able demonstrate that in a Schwarzschild black hole
infalling objects reach the singularity at the centre in a finite
time, and this fall is inevitable if the infalling object crosses the
horizon.
Penrose terms such surfaces that act as one way membranes as a trapped
surface. 

\begin{figure}[!t]
  \caption{Sir Roger Penrose is the Emeritus Rouse Ball Professor of
    Mathematics at the University of Oxford, an emeritus fellow of
    Wadham College, Oxford and an honorary fellow of St John's
    College, Cambridge, and of University College London (UCL).
    (Photo by Cirone-Musi, Festival della Scienza, CC BY-SA 2.0, https://commons.wikimedia.org/w/index.php?curid=19318743)}
  \includegraphics[width=2.5truein]{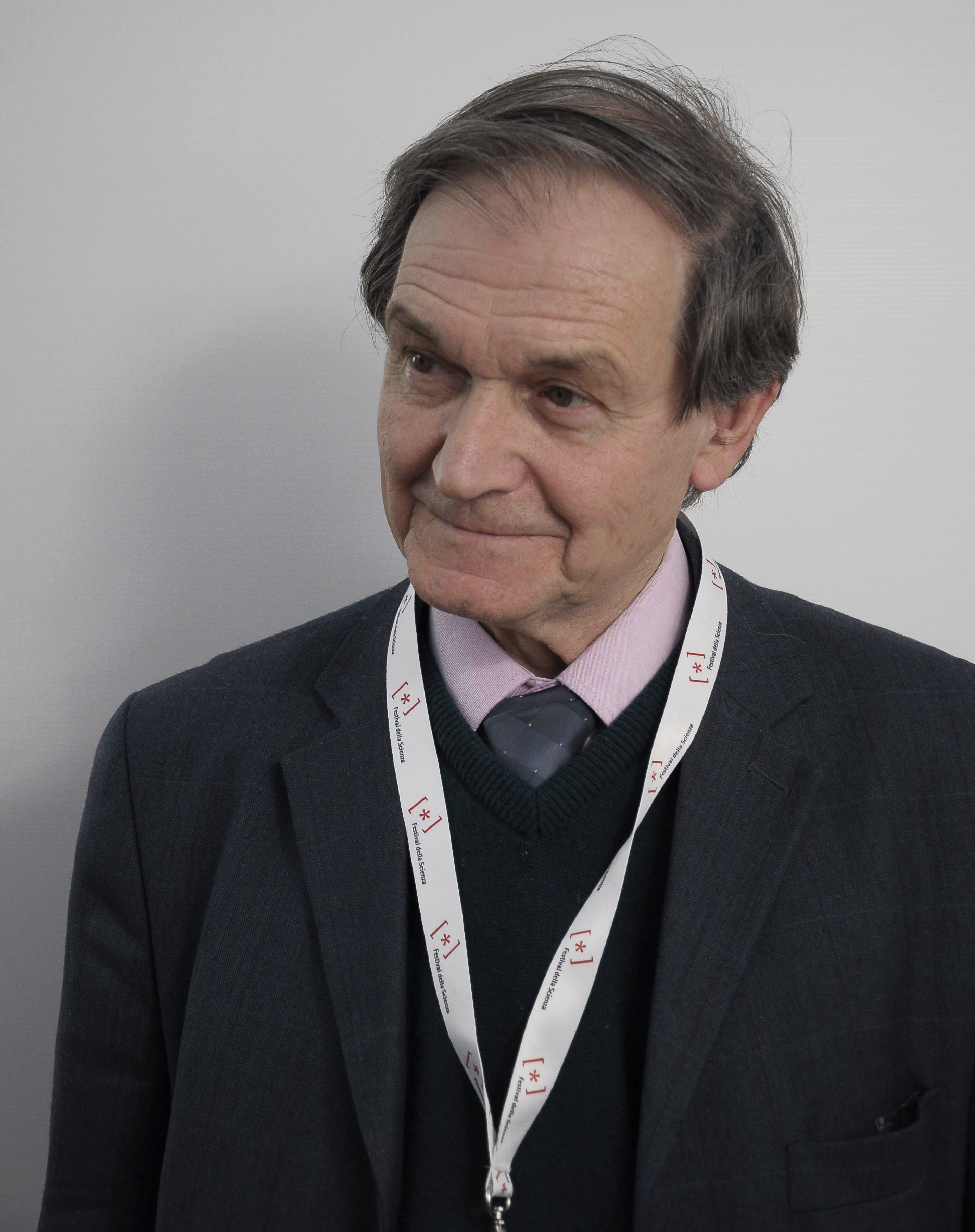} 
\end{figure}

Penrose proceeded further and was able to prove that if a trapped
surface forms during collapse of a star then the formation of a
singularity at the centre is inevitable.
This is the singularity theorem that he proved.
The proof required application of topological methods in general
relativity.
The implication of this theorem is that a star can collapse and form a
black hole, this astrophysical phenomena can lead to the formation of
a black hole.
Given that the known black hole solutions are stationary, this is a
significant step that connects these idealized solutions with complex
reality. 

This proof was timely as there were doubts about whether black holes
can form in astrophysical processes or not.
Measurement of redshift of quasars like 3C273 already implied that
the total emission from these was very large and accretion around
black holes appeared to be the only feasible explanation at the time.

The proof also brings out an internal limitation of the general theory
of relativity.
Existence of singularities implies that the theory breaks down at some
points and hence the theory has limitations.  
It is expected that as and when the quantum theory of gravity is
found, the singularities will be replaced by something more
tractable. 

Roy Kerr provided solution for a rotating black hole in 1963.
This is the black hole solution that is clearly relevant in
astrophysics as all stars have a non-zero angular momentum and we
expect black holes that form due to collapse to have some angular
momentum as well.

Roger Penrose proposed a process that can lead to extraction of energy
from a rotating black hole.
This is also a very significant contribution.

The singularity theorems were generalized by Hawking and Penrose to
prove that the Universe had a singularity in the past. 

However, what remains unproven is the so called {\sl cosmic censorship
  conjecture}.
The conjecture states that a space-time singularity is always
surrounded by a horizon and hence cannot be seen by a distant
observer.

\section{Super Massive Black Holes}

Evidence has grown over the last five decades for presence of super
massive black holes at the centre of each galaxies.
This evidence has come from a variety of observations of velocities in 
the vicinity of the black hole.
These observations are of stars or hot gas orbiting the black hole.
It is interesting that we had much better evidence for existence of a
super massive compact objects in other galaxies well before such
observations were attempted in our own galaxy.
A major obstacle in such observations in the Galaxy is scattering and
absorption by the intervening gas and dust, this effect is given the
name {\sl extinction} in astronomy. 
One half of the Nobel prize for physics has been given to Reinhard
Genzel and Andrea Mia Ghez for their work that has established the
presence of a super massive compact object at the centre of the
Galaxy.

\begin{figure}[!t]
  \caption{Reinhard Genzel  is a German astrophysicist, co-director of
    the Max Planck Institute for Extraterrestrial Physics, a professor
    at LMU and an emeritus professor at the University of California,
    Berkeley. 
    (Photo by MPE - Own work, CC BY-SA 3.0, https://commons.wikimedia.org/w/index.php?curid=18151111)}
  \includegraphics[width=2.5truein]{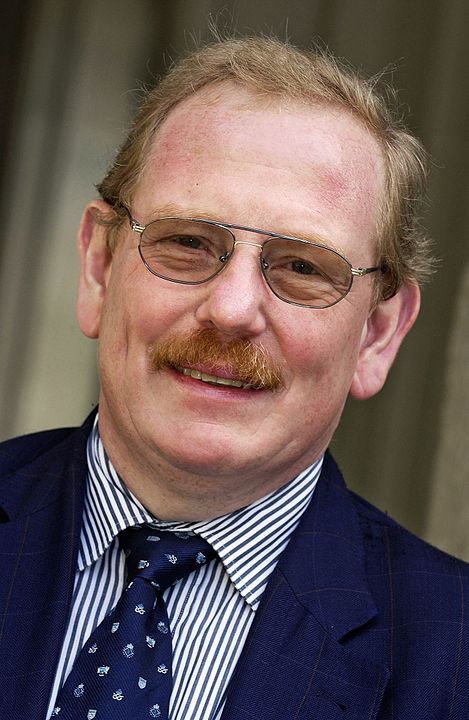} 
\end{figure}

The program to observe the central region of the Galaxy was started by
Reinhard Genzel nearly three decades ago.
Andrea Ghez started her own program a few years later.  
He identified a near infra-red band (K band, $\lambda \sim
2.2$~$\mu$m) as a suitable band for observations of the galactic
centre.
Extinction reduces the flux by nearly three orders of magnitudes in
the K band, the situation is much worse in the optical.
The first challenge that came up was the effect of the atmosphere.
Variations of density and temperature in the atmosphere lead to
distortion of the incoming wave front: astronomers refer to this
effect as {\sl seeing}. 
We are familiar with this effect in terms of twinkling of stars: the
image of the star shifts around by a small amount and its intensity
fluctuates.
In images of stars taken over a long period, we get the sum of all the
shifted positions and hence image of each star becomes a bit blurry. 
The number of stars in the direction of the galactic centre is very
large and therefore images of stars start to overlap.
The first technique used to get images without this problem was to
take many short images, align them and obtain a deep image of the
region without blurring of images.
These early studies allowed astronomers to get first estimates of the
speeds of stars in the central region and therefore constrain the mass
contained near the centre.

Shift to larger telescopes permitted astronomers to take deeper images
and observe fainter stars.
The most significant development was the use of adaptive optics.

\subsection{Active Optics and Adaptive Optics}

Active optics was developed three decades ago to overcome problems
that prevented construction of large telescopes.
Astronomers use reflector telescopes as mirrors can be supported from
the base and these do not suffer from chromatic aberrations.
As telescopes are reoriented to look at sources in different parts of
the sky, the shape of the mirror changes under its own weight.
In order to prevent this from causing distortion of images, thick
glass was used to make mirrors with the typical thickness being about
$1/5$ of the diameter of the mirror.
For large mirrors this causes two problems: the mirrors become very
heavy, and, the response of the material to changing temperature
causes distortion in shape as the temperature is not constant through
the body of the mirror.

Active optics was an attempt to work with thin mirrors of large
diameter where the shape is adjusted with the help of a computer model
that takes the orientation and temperature into account.
It is this technology that has allowed construction of telescopes with
large diameters in the last three decades and the same technology will
be used for making even larger telescopes with composite mirrors,
e.g., the thirty meter telescope (TMT) that India is also involved
in.

\begin{figure}[!t]
  \caption{Andrea Mia Ghez is an American astronomer and professor in
    the Department of Physics and Astronomy at the University of
    California, Los Angeles. Her research focuses on the center of the
    Milky Way galaxy.  
    (Photo by John D. - https://www.nsf.gov/discoveries/disc\_images.jsp?cntn\_id=133541\&amp;org=NSF, Public Domain, https://commons.wikimedia.org/w/index.php?curid=94809121)}
  \includegraphics[width=3truein]{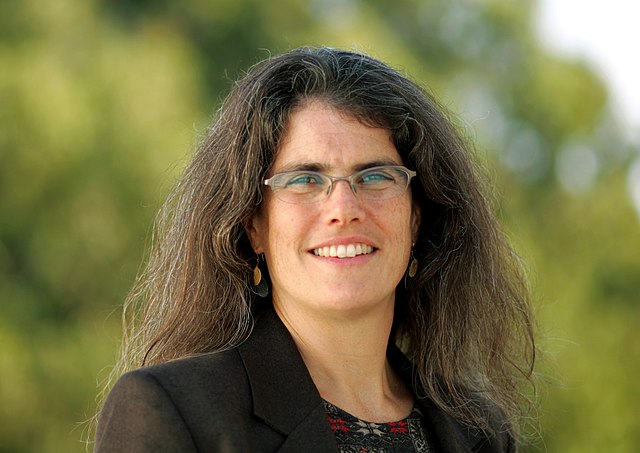} 
\end{figure}

Adaptive optics makes use of the technology developed for active
optics to adjust the shape of the mirror for distortion of the
incoming wave front by the atmosphere.
This requires a measurement of the incoming wavefront and a rapid
adjustment of the shape of the mirror.
Several approaches have been developed for this.
If we have a bright point source in the field then we can sense the
wavefront by an array of lenses/mirrors: if the wavefront did not have
any distortions then all images will form at the same relative
location, whereas there will be small shifts if the wavefront cannot
be described as a plane wave.
The amount of shift can be used to sense the level of distortion and
an appropriate correction can be applied.

Bright stars are not present in all directions and astronomers use
laser guide stars to overcome this problem.  
Adaptive optics permits astronomers to obtain diffraction limited
images and hence observe much fainter sources while improving
localization on the sky.

\subsection{Zooming in}

Observations with large telescopes: Keck telescopes in Hawaii and the
Very Large Telescopes (VLT) in Chile have allowed astronomers to
observe stars and track these in their orbits.
The orbital parameters of these stars constrain the mass contained
inside the orbit.
A key step was the discovery of a few stars with
time periods short enough for them to be tracked for more than
one complete orbit. This enables reliable determination of orbital
parameters for these stars.
The orbital parameters (semi-major axis and time period) can be used
to deduce the enclosed mass.  
A very important inference from these studies has been that the mass
contained within the orbit reaches a constant value, indicating that
the contribution in the innermost regions is from a single source and
individual stars do not contribute any significant mass.
Indeed, the contribution of stars to the total mass up to a
radius that is twenty thousand times the orbit of the Earth around
the Sun is insignificant, i.e., the mass of the compact object dominates
in this region. 
This, in essence is the observational evidence of the presence of an
object with a mass of about $4 \times 10^6$ times the mass of the Sun
in a region that is at most twenty times larger than the orbit of the
Earth around the Sun.
This is only about 200 times larger than the Schwarzschild
radius\cite{2020ApJ...889...61P}.
The mass remains almost constant to a scale that is a thousand times
larger than this. 

\subsection{Recent Developments}

Astronomers are now starting to use interferometry for observations of
the galactic centre.
This will improve the resolving power by an order of magnitude and
improve determination of orbital parameters.
This will also permit discovery of fainter objects.
Improved sensitivity is expected to permit testing general theory of
relativity via observations of stars in orbits around the compact
object. 

\section{Summary}

The Nobel prize for Physics in this year has recognized the
theoretical work that proved that it is possible for black holes to
form through astrophysical processes.
The work by Roger Penrose is built on foundations laid by a number of
scientists, most importantly Amal Kumar Raychaudhuri.
The Raychaudhuri equation was critical to the proof of the singularity
theorem given by Roger Penrose.

Recognition of the work by Reinhard Genzel and Andrea Mia Ghez in
establishing the presence of a super massive compact object at the
centre of the Galaxy stops short of calling it a black hole.
Observations using other means have established a stronger constraint
on supermassive compact objects in other galaxies.
Perhaps the observational discovery of black holes will be recognized
later in another Nobel.

Roy Kerr, who found the solution for a black hole with angular
momentum has also contributed significantly in this journey.
Without the Kerr solution, there would have been many concerns about
realistic models of black holes.

Readers can study more details at the Nobel foundation
website\footnote{https://www.nobelprize.org/uploads/2020/10/advanced-physicsprize2020.pdf}
and other reviews\cite{2020arXiv201106656S}.

I would like to thank Professor Rajaram Nityananda for useful comments
on an earlier draft of the manuscript.

\end{document}